\documentclass[aps,prl,floats,twocolumn,showpacs,superscriptaddress]{revtex4}

\usepackage{graphicx,epsfig}
\usepackage{times}
\usepackage{graphics,dcolumn,bm,fleqn,epic,eepic,float}
\usepackage{amssymb,amsmath,multirow,rotate,color}

\begin{document}

\title{Explosive Synchronization Transitions in Scale-free Networks}

\author{Jes{\'u}s G{\'o}mez-Garde\~{n}es}

\affiliation{Departamento de F{\'i}sica de la Materia Condensada,
  Universidad de Zaragoza, Zaragoza E-50009, Spain}

\affiliation{Institute for Biocomputation and Physics of Complex
Systems (BIFI), University of Zaragoza, Zaragoza 50009, Spain}

\author{Sergio G{\'o}mez}

\affiliation{Departament d'Enginyeria Inform{\`a}tica i
  Matem{\`a}tiques, Universitat Rovira i Virgili, 43007 Tarragona,
  Spain}

\author{Alex Arenas}

\affiliation{Institute for Biocomputation and Physics of Complex
Systems (BIFI), University of Zaragoza, Zaragoza 50009, Spain}

\affiliation{Departament d'Enginyeria Inform{\`a}tica i
  Matem{\`a}tiques, Universitat Rovira i Virgili, 43007 Tarragona,
  Spain}

\author{Yamir Moreno}

\affiliation{Institute for Biocomputation and Physics of Complex
Systems (BIFI), University of Zaragoza, Zaragoza 50009, Spain}

\affiliation{Departamento de F\'{\i}sica Te\'orica, Facultad de
  Ciencias, Universidad de Zaragoza, Zaragoza 50009, Spain}

\date{\today}

\begin{abstract}
The emergence of explosive collective phenomena has recently
attracted much attention due to the discovery of an
explosive percolation transition in complex networks. In this
Letter, we demonstrate how an explosive transition shows up in the
synchronization of complex heterogeneous networks by incorporating a
microscopic correlation between the structural and the dynamical
properties of the system. The characteristics of this explosive transition are
analytically studied in a star graph reproducing the results obtained
in synthetic scale-free networks. Our findings represent the first
abrupt synchronization transition in complex networks thus providing a
deeper understanding of the microscopic roots of explosive
critical phenomena.
\end{abstract}


\maketitle

Synchronization is one of the central phenomena representing the
emergence of collective behavior in natural and synthetic complex
systems \cite{winfree,strogatzsync,piko}. Synchronization processes
describe the coherent dynamics of a large ensemble of interconnected
autonomous dynamical units, such as neurons, fireflies or cardiac
pacemakers. The seminal works of Watts and Strogatz
\cite{watts,strogatz} pointed out the importance of the structure of
interactions between units in the emergence of synchronization, which gave rise
to the modern framework of complex networks \cite{report}.

Since then, the phase transition towards synchronization
has been widely studied by considering non-trivial networked
interaction patterns \cite{syncrep}. Recent results have shown that the
topological features of such networks strongly influence both the
value of the critical coupling, $\lambda_c$, for the onset of
synchronization \cite{MorenoPacheco,lee,arenas,zk,JGGPRL} and the
stability of the fully synchronized state
\cite{pecora,barahona,motter1,zhou}.
The case of scale-free (SF) networks has deserved special attention as they are
ubiquitously found to represent the
backbone of many complex systems. However, the topological
properties of the underlying network do not appear to affect the order of the
synchronization phase transition, whose second-order nature remains
unaltered \cite{MorenoPacheco}.

More recently, the study of explosive phase transitions in complex
networks has attracted a lot of attention since the discovery of an
abrupt percolation transition in random \cite{Science} and SF networks \cite{FortunatoPRL,ChoPRL}.
However, several questions about the
microscopic mechanisms responsible of such an explosive
percolation transition and their possible existence in other
dynamical contexts remain open. In this line, we conjecture that
such dynamical abrupt changes occur when both, the local heterogeneous structure
of networks and the dynamics
on top of it, are positively correlated.

In this Letter, we prove our conjecture in the context of the
synchronization of Kuramoto oscillators. We show that an explosive
synchronization transition emerges in SF networks when the natural
frequency of the dynamical units are positively correlated with the
degree of the units. Furthermore, we analytically study this
first-order transition in a star graph and show that the combination
of heterogeneity and the above correlation between structural and
dynamical features are at the core of the explosive synchronization
transition.


Let us consider an unweighted and undirected network of $N$ coupled
phase-oscillators. The phase of each oscillator, denoted by
$\theta_i(t)$ ($i=1,\ldots,N$), evolves in time according to the
Kuramoto model \cite{kuramoto75}:
\begin{equation}
\dot{\theta}_i=\omega_i + \lambda\sum_{j=1}^{N}
A_{ij}\sin(\theta_j-\theta_i),\;\;{\mbox{with}}\;i=1,...,N
\label{KM}
\end{equation}
where $\omega_i$ stands for the natural frequency of oscillator
$i$. The connections among oscillators are encoded in the adjacency
matrix of the network, ${\bf A}$, so that $A_{ij}=1$ when oscillators
$i$ and $j$ are connected while $A_{ij}=0$ otherwise. Finally, the
parameter $\lambda$ accounts for the strength of the coupling among
interconnected nodes.

The original Kuramoto model assumed that the oscillators were
connected all-to-all, {\em i.e.} $A_{ij}=1$ $\forall i\neq j$. In this
setting, a synchronized state, {\em i.e.} a state in
which $\dot{\theta_{i}}(t)=\dot{\theta_{j}}(t)$ $\forall i,j$ and
$\forall t$, shows up when the strength of the coupling $\lambda$ is
larger than a critical value \cite{kuramoto75,kurabook,conradrev}. To monitor such
synchronization transition as $\lambda$ grows, the following complex order parameter,
which quantifies the degree of synchronization
among the $N$ oscillators, is used \cite{strogatzkura}:
\begin{equation}
r(t)e^{i\Psi(t)}=\frac{1}{N}\sum_{j=1}^{N} e^{i\theta_j(t)}
 \label{r_kura}
\end{equation}
The modulus of the above order parameter, $r(t)\in [0,1]$, measures
the coherence of the collective motion, reaching the value $r=1$ when the system is
fully synchronized, while $r=0$ for the incoherent solution.
On the other hand, the value of $\Psi(t)$ accounts of the average
phase of the collective dynamics of the system. Typically, the average
(over long enough times) value of $r$ as a function of the coupling
strength $\lambda$ displays a second-order phase transition from $r=0$
to $r=1$ with a critical coupling $\lambda_c=2/(\pi g(\omega=0))$, where
$g(\omega)$ is the distribution of the natural frequencies,
$\{\omega_{i}\}$, and it is assumed to be unimodal and even
\cite{strogatzkura}.


Here we will focus on the influence of the dynamical and topological
characteristics at the local level (the nodes of the network and their
interactions) in the emergence of global synchronization. In
particular, we will identify the internal frequency of each node $i$ directly with
its degree $k_{i}$, so that
$\omega_i=k_i$ in Eqs. (\ref{KM}). Note that this prescription automatically
sets that the distribution of frequencies $g(\omega) = P(k)$ but not vice versa \cite{note}.

\begin{figure}[!t]
\centering
\includegraphics[width=3.3in]{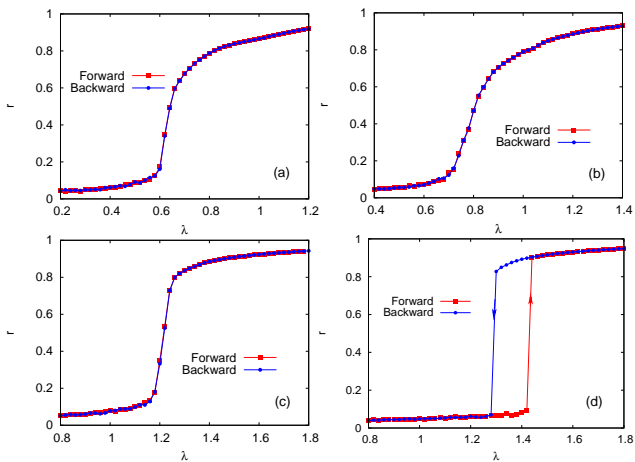}
\caption{(color online) Synchronization diagrams $r(\lambda)$
  for different networks constructed using the interpolation model introduced in
  \cite{BAER}. The $\alpha$ values in each panel are (a) $\alpha=1$
  (ER), (b) $\alpha=0.6$, (c) $\alpha=0.2$ and (d) $\alpha=0$
  (BA). The four panels show both {\em Forward} and {\em Backward}
  continuations in $\lambda$ using increments of
  $\delta\lambda=0.02$. The size of the networks is $N=10^3$ and the
  average degree is $\langle k\rangle=6$.}
\label{fig1}
\end{figure}

To study the effects of the correlation between dynamical and
structural attributes, we simulate the Kuramoto model on top of a
family of networks generated according to \cite{BAER}. This model allows to
construct networks with the same average connectivity, $\langle k
\rangle$, interpolating from Erd\"os-R\`enyi (ER) graphs to
Barab\`asi-Albert (BA) SF networks by tuning a single parameter
$\alpha$. The growth of the networks assumes that a newly added node either
attaches randomly with probability
$\alpha$ or preferentially to those nodes with large degree with
probability $(1-\alpha)$. In this way, $\alpha=1$ gives
rise to ER graphs with a Poissonian degree distribution whereas for
$\alpha=0$ the resulting networks are SF with $P(k)\sim k^{-3}$.
Intermediate values $\alpha\in(0,1)$ tune the
heterogeneity of the network, which increases when going from
$\alpha=1$ to $\alpha=0$. In the four panels of Fig.~\ref{fig1}, we
report the synchronization diagrams of four network topologies
constructed using this model. The limiting cases of ER and BA networks
correspond to panel \ref{fig1}a and \ref{fig1}d, respectively. The size of these networks
are $N=10^3$ while the average connectivity is set to $\langle k\rangle=6$.

For each panel in Fig.~\ref{fig1} we have computed two synchronization
diagrams, $r(\lambda)$, labeled as {\em Forward} and {\em Backward}
continuations. The former diagram is computed by increasing
progressively the value of $\lambda$ and computing the stationary
value of the order parameter $r$ for $\lambda_0$,
$\lambda_{0}+\delta\lambda$,...,
$\lambda_{0}+n\delta\lambda$. Alternatively, the backward continuation
is performed by decreasing the values of $\lambda$ from
$\lambda_{0}+n\delta\lambda$ to $\lambda_0$. The panels \ref{fig1}a,
\ref{fig1}b and \ref{fig1}c show a typical second-order transition
with a perfect match between the backward and forward synchronization
diagrams. Importantly, the onset of synchronization is delayed as the
heterogeneity of the underlying graph (and thus that of the frequency
distribution $g(\omega)$) increases.

The most striking result is however observed for the BA network (panel
\ref{fig1}d) in which a sharp, first-order synchronization transition
appears. In the case of the forward continuation diagram the order
parameter remains $r\simeq 0$ until the onset of synchronization in
which $r$ jumps suddenly to $r\simeq 1$ pointing out that almost all
the network has reached the synchronous motion. Moreover, the diagram
corresponding to the backward continuation also shows a sharp
transition from the fully synchronized state to the incoherent
one. The two sharp transitions takes place at different values of $r$
so that the whole synchronization diagram displays a strong
hysteresis.

\begin{figure}[!t]
  \begin{center}
  \begin{tabular}[t]{c}
    \mbox{\includegraphics[height=2.50in,angle=-0]{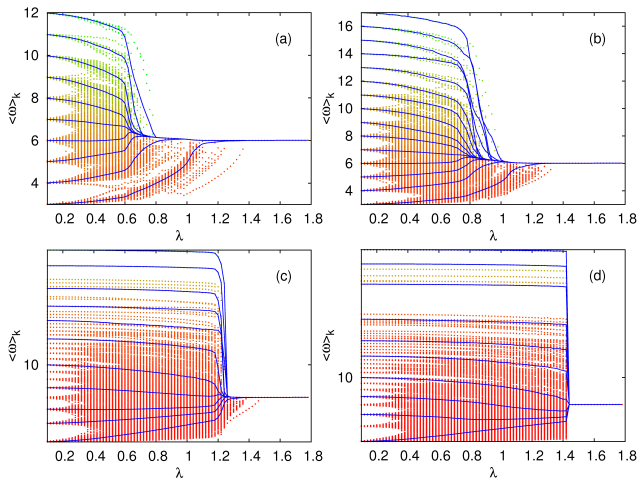}}
  \end{tabular}
  \end{center}
\caption{(color online) The panels show the evolution of the effective frequencies of
  the nodes along the (forward) continuation in the four model
  networks of Fig.~\ref{fig1}. The colored dots account for single-node
  values (colors stand for their respective degree) while the solid lines
  show the evolution of the average value of the effective
  frequencies of nodes having the same degree.}
\label{fig2}
\end{figure}

To analyze deeply the change of the order of the
synchronization transition, we have monitored the
evolution of the dynamics for every node by computing their
effective frequency along the forward continuation, see Fig.~\ref{fig2} . The effective
frequency of a node $i$ is defined as
\begin{equation}
\omega^{\mbox{\scriptsize eff}}_{i}=\frac{1}{T}\int_{t}^{t+T}\dot{\theta_{i}}(\tau)\;{\mbox
    d}{\tau}\;,
\end{equation}
with $T\gg 1$. We have also computed the evolution of
$\omega^{\mbox{\scriptsize eff}}_{i}$ within a degree class $k$,
$\langle\omega\rangle_{k}$, averaging over nodes having identical
degree $k$:
\begin{equation}
\langle \omega\rangle_{k}=\frac{1}{N_{k}}\sum_{[i|k_{i}=k]}\omega^{\mbox{\scriptsize eff}}_{i}\;,
\end{equation}
where $N_{k}=NP(k)$ is the number of nodes with degree $k$ in the
network. From the panels in Fig.~\ref{fig2} we observe that the
individual frequencies and the different curves
$\langle\omega\rangle_{k}(\lambda)$ converge progressively to the
average frequency of the system $\Omega=\langle k\rangle=6$ until full
synchronization is achieved. Panel \ref{fig2}a (ER graph) shows that
the convergence to $\Omega$ is first achieved by those nodes with
large degree while the small $k$-classes achieve full synchronization
later on. As the heterogeneity of the network increases (see
$\alpha=0.6$ and $\alpha=0.2$ in panels \ref{fig2}b and \ref{fig2}c,
respectively) the differences in the convergence of the $k$-classes
decrease. Finally, for the BA network (Fig.~\ref{fig2}d), we observe
that nodes (and thus the different $k$-classes) retain their natural
frequencies until they become almost fully locked, which signal the
abrupt synchronization observed in Fig.~\ref{fig1}d. Thus, the
first-order transition of the BA network corresponds to a process in
which no microscopic signals of synchronization are observed until the
critical coupling $\lambda_c$ is reached.

\begin{figure}[!t]
\begin{center}
\includegraphics[width=3.4in,angle=0]{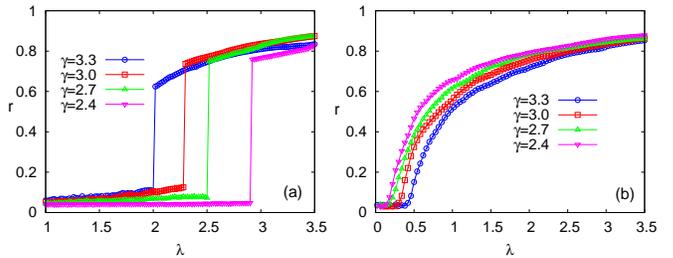}
\end{center}
\caption{(color online) Panel (a) shows the synchronization diagrams $r(\lambda)$ for
  several SF networks constructed via the configurational model. All
  the networks have a degree-distribution $P(k)\sim k^{-\gamma}$ with
  $\gamma=2.4$, $2.7$, $3.0$, and $3.3$ while $N=10^3$. The steps of the
  continuation are set to $\delta \lambda=0.02$. In panel (b) we show
  the synchronization diagrams of the same SF networks without the
  local correlation between degrees and natural frequencies, {\em
    i.e.} $\omega_{i}\neq k_{i}$, while the distribution of natural
  frequencies is still $P(\omega)\sim\omega^{-\gamma}$.}
\label{fig3}
\end{figure}

To further explore the correspondence of the explosive synchronization
transition with the SF nature of the underlying graph, in
Fig.~\ref{fig3}.a we show the synchronization diagrams for different
uncorrelated SF graphs with different degree distribution' exponents. These
graphs have been constructed using the configurational model
\cite{conf} by imposing a degree distribution $P(k)\sim k^{-\gamma}$ with
$\gamma=2.4$, $2.7$, $3.0$ and $3.3$. The synchronization
diagrams are obtained by forward continuation
(as described above) starting at $\lambda=1$ and performing adiabatic
increments of $\delta \lambda=0.02$. Again, for each value of
$\lambda$ the Kuramoto dynamics is run until the value of $r$ reaches
its stationary state. From the figure it is clear that a first-order
synchronization transition appears for all the reported values of
$\gamma$ pointing out the ubiquity of the explosive synchronization
transition in SF networks. Moreover, the onset of synchronization,
$\lambda_c$, is delayed as $\gamma$ decreases, {\em i.e.} when the
heterogeneity of the graph increases.

Up to now, we have shown that the explosive synchronization transition
appears in SF when the natural frequencies of the nodes are correlated
with their degrees. To show that this correlation is the responsible
of such explosive transition, in Fig.~\ref{fig3}b we show the
synchronization diagram for the same SF networks used in
Fig.~\ref{fig3}a, but when the correlation between dynamics and
structure is broken in such a way that the same distribution for the
internal frequencies, $g(\omega)=\omega^{-\gamma}$ is kept. To this
end, we made a random assignment of frequencies to nodes according to
$g(\omega)$.  The plots reveal that now all the transitions turn to be
of second-order, thus recovering the usual picture of synchronization
phenomena in complex networks. Therefore, the first-order transition
arises due to the positive correlation between natural frequencies and
the degrees of the nodes in SF networks \cite{check}.

\begin{figure}[!t]
\includegraphics[width=3.4in]{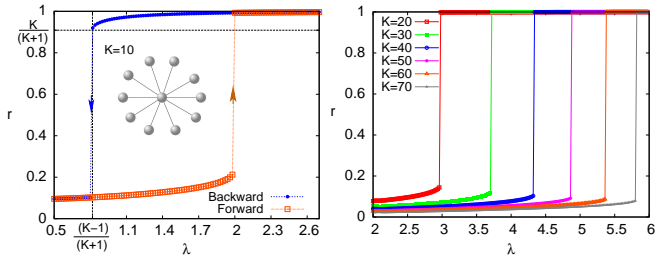}
\caption{(color online) We show the synchronization diagrams for the
  star graph [see the inset in plot (a)]. In (a) we show the (forward
  and backward) continuation diagrams for the case $K=10$ while (b)
  shows the forward continuation diagrams for different star graphs of
  different sizes, corresponding to $K=20$, $30$, $40$, $50$, $60$ and
  $70$.}
\label{fig4}
\end{figure}

All the simulations results presented corroborate our conjecture about
the explosive percolation transition in SF networks. To get analytical
insights, we reduce the problem studied to the analysis of the {\em
  star} configuration, a special structure that grasp the main
property of SF networks, namely the role of hubs.  Therefore, we
explore the synchronization transition of such a configuration and
show that it is indeed explosive when the correlation $\omega_i=k_i$
holds. A star graph (as shown in the inset of Fig.~\ref{fig4}a) is
composed by a central node (the hub) and $K$ peripheral nodes (or
leaves). Each of the peripheral nodes connects solely to the
hub. Thus, the connectivity of the leaves is $k_{i}=1$ ($i=1,...,K$)
while that of the hub is $k_{h}=K$. Let us suppose that the hub has a
frequency $\omega_{h}$ while all the leaves beat at the same frequency
$\omega$.

First we set a reference frame rotating with the average phase of the
system, $\Psi(t)=\Psi(0)+\Omega t$, being $\Omega$ the average
frequency of the oscillators in the star,
$\Omega=(K\omega+\omega_{h})/(K+1)$. In the following we set
$\Psi(0)=0$ without loss of generality so that the transformed
variables are defined as $\phi_{h}=\theta_{h}-\Omega t$ for the hub
and $\phi_{j}=\theta_{j}-\Omega t$ (with $j=1,...,K$) for the leaves.
Thus, the equations of motion for the hub and the leaves read:
\begin{eqnarray}
\dot{\phi_h}&=&(\omega_{h}-\Omega)+\lambda\sum_{j=1}^{K}\sin(\phi_{j}-\phi_{h}),
\label{eq:hub}
\\ \dot{\phi_j}&=&(\omega-\Omega)+\lambda\sin(\phi_{h}-\phi_{j}),\;{\mbox{with}}\;j=1...K.
\label{eq:leaves}
\end{eqnarray}
In this rotating frame the motion of the hub, Eq. (\ref{eq:hub}), can
be expressed as:
\begin{equation}
\dot{\phi_h}=(\omega_{h}-\Omega)+\lambda(K+1)r\sin(\phi_{h})\;,
\label{eq:hub2}
\end{equation}
note that in this new frame it is easy to identify that the dynamics
of the hub is governed by its new inherent frequency and the
superposition of a set of identical signals from the leaves. Now,
imposing that the phase of the hub is locked, $\dot{\phi_h}=0$, we
obtain:
\begin{equation}
\sin\phi_h=\frac{(\omega_{h}-\Omega)}{\lambda(K+1)r}\;.
\label{eq:sinphih}
\end{equation}

Now we consider the equations for the leaves, Eq. (\ref{eq:leaves}),
and evaluate the expression for $\cos\phi_{j}$ in the locked regime,
$\dot{\phi}_{j}=0$. After some algebra, we obtain the following
expression:
\begin{equation}
\cos\phi_{j}=\frac{(\Omega-\omega)\sin\phi_{h}\pm\sqrt{\left[1-\sin^2\phi_{h}\right]\left[\lambda^2-(\Omega-\omega)^2\right]}}{\lambda}\,.
\label{eq:cosphij}
\end{equation}
The above expression is valid only when
$(\Omega-\omega)\leq\lambda$. From this inequality we obtain the value
of the coupling $\lambda$ for which the phase-locking is lost, {\em
  i.e.} the critical coupling $\lambda_c=\Omega-\omega$. In our case,
we have $\omega_h=K$, $\omega=1$ and $\Omega=2K/(K+1)$ so that we
obtain a critical coupling $\lambda_c=(K-1)/(K+1)$. On the other hand,
we can derive the value $r_c$ of the order parameter at the critical
point by using Eq.~(\ref{eq:sinphih}) and
Eq.~(\ref{eq:cosphij}) to compute $r=\langle \cos(\phi)\rangle$ at
$\lambda_c$:
\begin{equation}
r_c=\frac{\cos(\phi_h)+K\cos(\phi_{j})}{K+1}\biggl\lvert_{\lambda_c}\biggr.=\frac{K}{(K+1)}\;.
\end{equation}
Therefore, as $r_c>0$, when the synchronization is lost there is a gap
in the synchronization diagram pointing out the existence of a
first-order synchronization transition. Moreover, as $K$ increases both
$\lambda_c$ and $r_c$ tend to $1$ thus confirming the first-order
nature of the transition in the thermodynamic limit,
$K\rightarrow\infty$. As shown in Fig.~\ref{fig4}a for the case
$K=10$, the theoretical values of $\lambda_c$ and $r_c$, are in
perfect agreement with results from numerical simulations. Finally, as
shown in Fig.~\ref{fig4}b, the stability of the unlocked phase regime,
$r\simeq 0$, increases with $K$ so that we can reach larger values of
$\lambda$ by continuing (forward) the solution with $r\simeq 0$. As a
result, this robustness of the incoherent solution leads to a larger
hysteresis cycle of the synchronization diagram as $K$ increases.


Summing up, we have shown that an explosive synchronization transition
occurs in SF networks when there is a positive correlation between the
structural (the degrees) and dynamical (natural frequencies)
properties of the nodes. This constitutes the first example of an
explosive synchronization transition in complex networks.
Moreover, we have shown that the emergence of such transition is
intrinsically due to the interplay between the local structure and
the internal dynamics of nodes rather than being caused by any
particular form of the distribution of natural frequencies. Our
findings provide with an explosive phase transition of an important
macroscopic phenomena, synchronization, in a widely studied dynamical
framework, the Kuramoto model, thus shedding light to the microscopic
roots behind these phenomena and paving the way to their study in
other dynamical contexts.

\acknowledgments J.G.-G. is supported by the MICINN through the
Ram\'on y Cajal Program. This work has been partially supported by the
Spanish DGICYT Grants FIS2008-01240, MTM2009-13848,
FIS2009-13364-C02-01, FIS2009-13730-C02-02, and the Generalitat de
Catalunya 2009-SGR-838.


\newpage

\begin{figure*}[!t]
\includegraphics[width=3.3in,angle=-90]{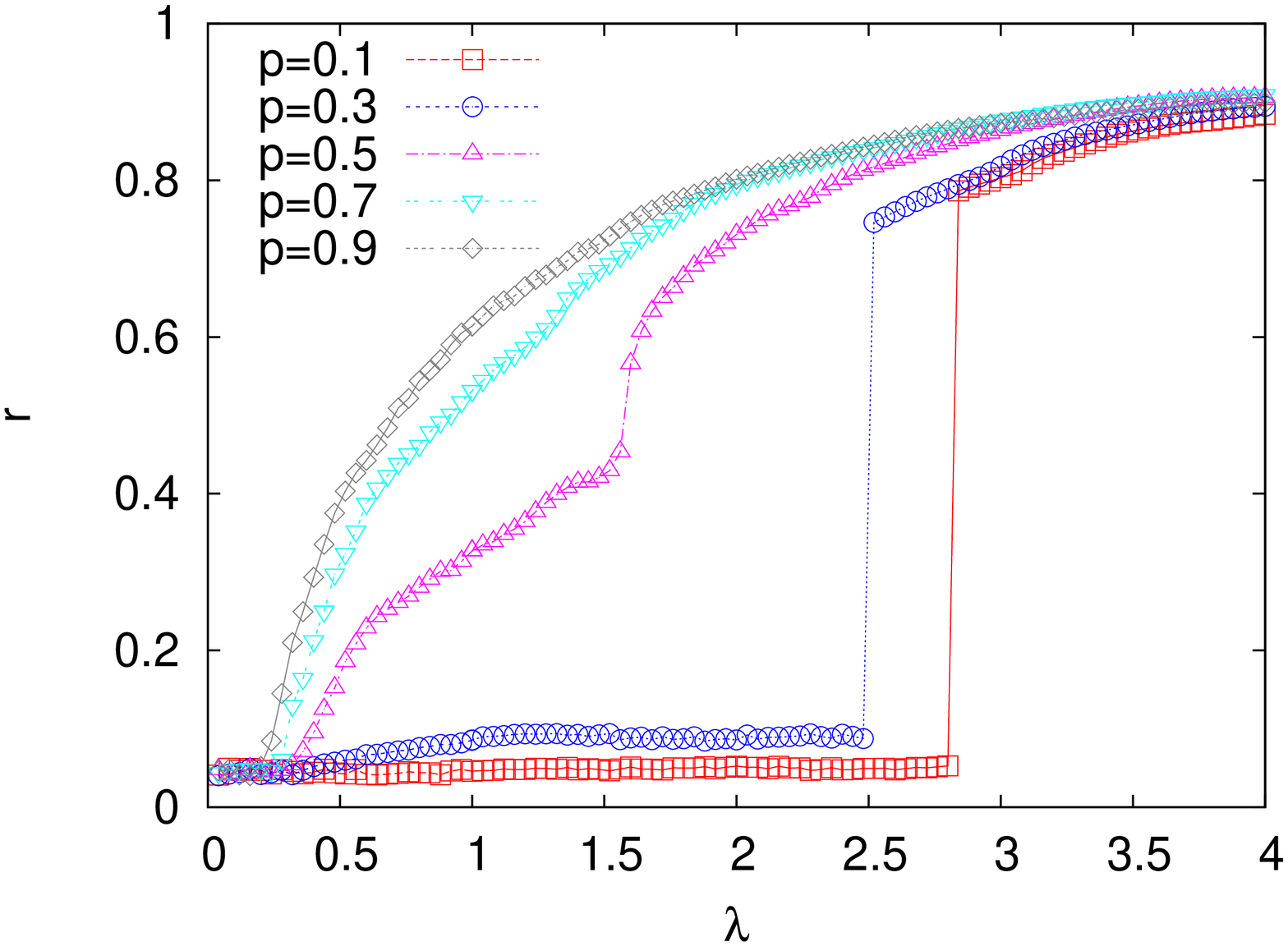}
\caption{Synchronization diagrams for intermediate degree-frequency
  correlation (see \cite{check} for further details).}
\label{figs}
\end{figure*}


\begin{thebibliography}{1}
\bibitem{note} It is worth stressing that our main results hold
when $w_i\sim k_i^{\alpha}$ with $\alpha> 0$, i.e.\ when $w_i$ is
positively correlated with $k_i$.
\end{thebibliography}
\end{document}